%% file: weakly_bioseg.tex
\documentclass{article}
\usepackage{spconf,amsmath,graphicx}

\usepackage{enumitem}
\setlist{nosep, leftmargin=14pt}

\usepackage{mwe} 
\usepackage{multirow}
\usepackage{booktabs}
\usepackage{amsmath}
\usepackage{amssymb}
\usepackage{comment}

\usepackage{hyperref}
\usepackage{xcolor}

\title{Weakly Supervised Nuclei Segmentation via Instance Learning}
\name{Weizhen Liu$^{\star,1}$, Qian He$^{\star,1,2,3}$, Xuming He$^{\dagger,1,4}$\thanks{$^{\star}$ denotes equal contribution. $^{\dagger}$ denotes corresponding author.}}
\address{$^{1}$ School of Information Science and Technology, ShanghaiTech University \\
$^{2}$ Shanghai Institute of Microsystem and Information Technology, Chinese Academy of Sciences\\
$^{3}$ University of Chinese Academy of Sciences\\
$^{4}$ Shanghai Engineering Research Center of Intelligent Vision and Imaging}
\begin{document}
%
\maketitle
\begin{abstract}
Weakly supervised nuclei segmentation is a critical problem for pathological image analysis and greatly benefits the community due to the significant reduction of labeling cost. Adopting point annotations, previous methods mostly rely on less expressive representations for nuclei instances and thus have difficulty in handling crowded nuclei. In this paper, we propose to decouple weakly supervised semantic and instance segmentation in order to enable more effective subtask learning and to promote instance-aware representation learning. To achieve this, we design a modular deep network with two branches: a semantic proposal network and an instance encoding network, which are trained in a two-stage manner with an instance-sensitive loss. Empirical results show that our approach achieves the state-of-the-art performance on two public benchmarks of pathological images from different types of organs. Our code is available at \href{https://github.com/weizhenFrank/WeakNucleiSeg}{https://github.com/weizhenFrank/WeakNucleiSeg}.
\end{abstract}

\graphicspath{ {./tex/imgs/} }

\input{tex/intro.tex}
\input{tex/method.tex}
\input{tex/exp.tex}

\input{tex/conclusion.tex}
\section{Compliance with Ethical Standards}
\label{sec:ces}
This research study was conducted retrospectively using human subject data made available in open access by \cite{kumar2017dataset} 
and \cite{naylor2018segmentation}. Ethical approval was not required as confirmed by the license attached with the open access data.

\section{Acknowledgments}
\label{sec:acknowledgments}
This work was supported by Shanghai Science and Technology Program 21010502700.

%


\small
\bibliographystyle{IEEEbib}
\bibliography{strings,refs}

\end{document}

%% file: tex/intro.tex
\begin{figure*}[t!]
		\centering
		\centerline{\includegraphics[width=0.82\textwidth]{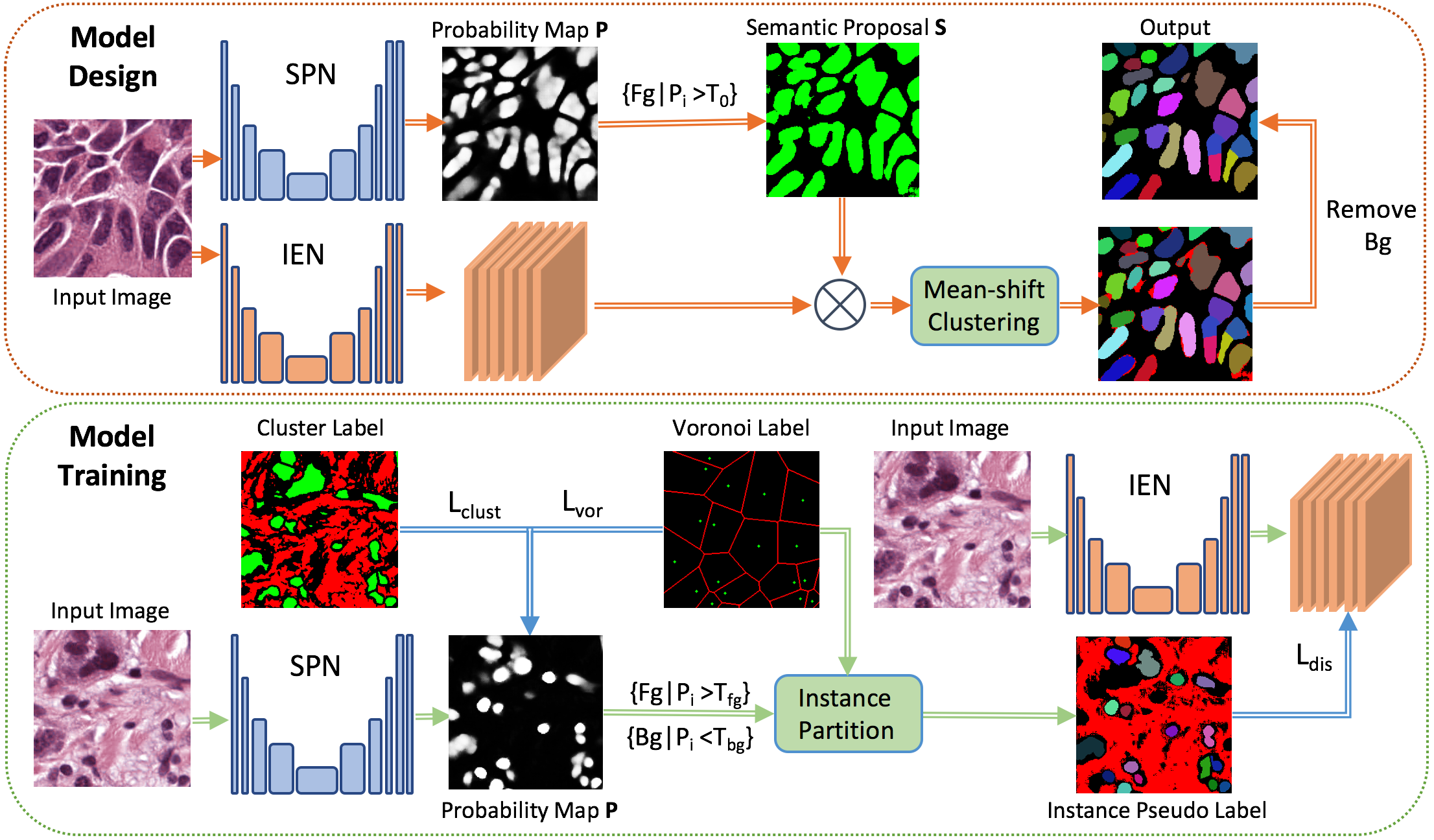}}
	%
	%
	\vspace{-0.3cm}
	\caption{Overview of our method. \textbf{Model Design}: Our model consists of two branches: a Semantic Proposal Network (SPN) to generate semantic proposals, and an Instance Encoding Network (IEN) to produce pixel-wise instance representations. We conduct instance grouping by mean-shift clustering on instance features selected by semantic proposals. \textbf{Model Training}: We train our model in a two-stage manner: We first generate cluster label and Voronoi label to train our SPN, and then generate instance pseudo labels by utilizing Voronoi partitions to divide predicted masks by the trained SPN, to train our IEN with a discriminative loss.}
	\label{fig:res}
	\vspace{-0.3cm}
\end{figure*}

\section{Introduction}
\label{sec:intro}

Nuclei segmentation is a crucial step in pathological image analysis and of great importance in clinical practice. Thanks to deep learning techniques, automated nuclei segmentation has made rapid progress recently.  
Nevertheless, most supervised methods require full-mask annotations, which are costly to obtain and hence largely deter their broad application in real-world scenarios. 
To alleviate this problem, recent efforts start to adopt weakly supervised learning strategy as a promising solution.

A commonly-adopted weak supervision for nuclei segmentation is based on central point annotation due to the dense distribution and semi-regular shape of nuclei. In particular, Qu et al.~\cite{qu2019weakly,qu2020weakly} first use the point annotation to generate pseudo pixel-level labels for the training of segmentation network with a CRF regularization loss. 
Other works utilize the Sobel filter to obtain pseudo edge maps for regularization in training~\cite{yoo2019pseudoedgenet,tian2020weakly}, or exploits objectness to generate pseudo mask for supervision~\cite{li2021point}. 
Those methods, however, typically share semantic and instance representations at the pixel level, and thus have difficulty in handling crowded nuclei instances due to the lack of expressive instance-aware representations. A few recent studies attempt to improve the feature learning by training a model to predict a peak-shape probability map centered at each nucleus~\cite{chamanzar2020weakly,dong2020towards}, or resort to cell/nuclei centroid detection and post-processing with Graph-Cut~\cite{nishimura2019weakly} or watershed~\cite{hou2019robust}. Such a loss design, nonetheless, often suffers from inaccurate instance boundaries due to the relatively weak supervision around the boundary regions. 
\cite{liu2020unsupervised,liu2020pdam,hsu2021darcnn,yang2021minimizing} tackle instance learning with limited annotations for domain adaptation, while their methods still require expensive fully labeled source data to help generate pseudo labels with high-quality boundaries for target data.

In this work, we propose an instance-aware learning strategy for weakly supervised nuclei segmentation in order to address the aforementioned limitations. Our main idea consists of two aspects: First, we decouple the semantic and instance segmentation process, enabling the model to effectively learn each of two subtasks as well as their fusion. Moreover, we introduce a discriminative loss to promote learning an instance-aware representation for nuclei pixels.   

To achieve this, we design a modular deep neural network consisting of two branches: a Semantic Proposal Network (SPN) to generate proposals for foreground and background masks, and an Instance Encoding Network (IEN) to perform instance feature embedding for  the subsequent grouping of the foreground proposal to produce final segmentation results. 
To train our model with point supervision, we develop a two-stage 
training strategy that sequentially learns the semantic and instance representation. Specifically, inspired by~\cite{qu2019weakly}, we first generate foreground pseudo labels with an adaptive k-means clustering and then train the SPN module. Subsequently, we use the predicted foreground masks from the SPN and Voronoi partition to obtain pseudo labels for nuclei instances. We finally learn the feature embedding network of our IEN module with an instance-level discriminative loss~\cite{de2017semantic}.

We evaluate our method on two public benchmarks, i.e., MultiOrgan~\cite{kumar2017dataset} and TNBC~\cite{naylor2018segmentation}, and the empirical results show that our method achieves state-of-the-art performance on both benchmarks.

\vspace{-0.3cm}

%% file: tex/method.tex
\section{Method}
\label{sec:method}
In this section, we will describe our method for weakly supervised nuclei segmentation. 
Below we first introduce our model design and then explain model training in detail.

\subsection{Model Design}
Given an input image $\mathbf{I} \in \mathbb{R}^{H \times W \times 3}$, we aims to generate its semantic mask $\mathbf{M} \in \{0,1\}^{H \times W}$ for foreground nuclei and background tissue, as well as foreground nuclei instance partition. To this end, we decouple semantic and instance segmentation process, and propose a model consisting of two main modules: a Semantic Proposal Network (SPN) and an Instance Encoding Network (IEN). 
Our SPN takes an input image $\mathbf{I}$ and generates a semantic probability map $\mathbf{P}\in [0,1]^{H \times W}$ for foreground nuclei and background tissue. 
Our IEN takes the same image $\mathbf{I}$ as input, and generates instance feature representation $\mathbf{F} \in \mathbb{R}^{H \times W \times D}$. 

During model inference, we first generate a foreground candidate mask  $\mathbf{S}$ by thresholding the output $\mathbf{P}$ of SPN with a relatively low value $T_0$, which gives better foreground recall. We then use the mask $\mathbf{S}$ to select the instance features from $\mathbf{F}$, and employ the mean-shift clustering to produce instance candidates. We finally remove the background cluster that typically has the largest number of connected components. An overview of our model architecture is shown in Fig.~\ref{fig:res}.

We instantiate both SPN and IEN networks with a UNet-like architecture. Specifically, we use an $1\times 1$ conv in the last layer for pixel-wise classification in the SPN and for feature embedding in the IEN. Additionally, for the IEN, we add a CoordConv~\cite{liu2018intriguing} layer in the beginning to combine color and position information for each pixel. 
\vspace{-0.3cm}

\subsection{Model Training}
We train our model in two stages, which starts from the semantic proposal network and then learns the instance grouping network. Both stages first generate pseudo labels and then train the networks with task-specific losses, which are introduced below for each stage respectively.  

\vspace{1mm}\noindent
\textbf{Training of SPN:} 
To generate semantic pseudo labels from point annotations, we adopt the k-means clustering and Voronoi partition strategy as in~\cite{qu2019weakly}. In order to cope with varying imaging conditions, we develop an image-adaptive feature representation for the k-means algorithm.  
Specifically, for pixel $i$ in an input RGB image $\mathbf{I}_k$, $c_i = \left(r_{i}, g_{i}, b_{i}\right)$ denotes its RGB value and $d_i$ represents the distance from pixel $i$ to its nearest point label truncated by $d^*$. We design the feature vector used for clustering as $f_{i} = (d_{i}, \hat{r}_{i}, \hat{g}_{i}, \hat{b}_{i})$, where $(\hat{r}_{i}, \hat{g}_{i}, \hat{b}_{i}) = \lambda * (r_{i}, g_{i}, b_{i}) / \sigma_k$. Here instead of using a global scaling term $\lambda$ to balance color and distance, we compute an image-specific $\sigma_k$ to scale RGB value so that our algorithm can adapt to the color distribution of each image. To compute $\sigma_k$, we first collect all  $L\times L$ patches centered around central point labels, and then compute the color distance of each pixel in a patch to its $9 \times 9$ neighbor pixels. $\sigma_k$ is set as the standard deviation of all computed color distance values within this image. 

Resorting to the k-means clustering and Voronoi partition, we generate cluster labels and Voronoi labels as shown in Fig.~\ref{fig:res}, with green denoting foreground, red denoting background, and black denoting unlabeled pixels. 
Given those pseudo labels, we train the SPN module with cross-entropy loss on the partially labeled pixels.

\vspace{1mm}\noindent
\textbf{Training of IEN:}
We introduce a discriminative loss~\cite{de2017semantic} to train our IEN in order to learn an instance-aware representation for nuclei pixels. 
To generate instance pseudo labels, we first generate a foreground mask by thresholding the output of the trained SPN with a probability value $T_{fg}$, and then utilize Voronoi partitions to divide it into nuclei instance labels. In addition, we treat the background region as a background instance and generate its label by selecting pixels with foreground probability lower than $T_{bg}$. 

Our discriminative loss $L_{dis}$ for each image consists of three terms: an intra-class term $L_{intra}$ to pull all pixel embeddings within an instance towards its mean embedding, an inter-class term $L_{inter}$ to push the mean embeddings of all instances away from each other, and a regularization term to pull all instances towards the origin. We adopt the hinge losses as in~\cite{de2017semantic}, which are denoted as  $h_{+}(a, b) = \max(0, ||a-b|| - \delta_{intra})$ and $h_{-}(a, b) = \max(0, \delta_{inter} - ||a-b||)$. Given $C$ instances from the pseudo label generation, the loss function for an image can be written as, 
\begin{align}
L_{dis} &= \frac{1}{C}L_{intra} + \frac{1}{C(C-1)} L_{inter} +  \frac{\gamma}{C}\sum_{c=1}^C ||\mu_c||\\
L_{intra} &= \sum_{c\neq bg} \frac{1}{N_c} \sum_{i=1}^{N_c} h_{+}^2 (\mu_c,x_i)  + \frac{\alpha } {N_{bg}}\sum_{i=1}^{N_{bg}} h_{+}^2 (\mu_{bg}, x_i)\nonumber \\
L_{inter} &= \underset{c_{A} \neq c_{B} \neq bg}{\sum_{c_{A}} \sum_{c_{B}}} h_{-}^2 (\mu_{c_{A}},\mu_{c_{B}})+ 2\alpha\sum_{c\neq bg} h_{-}^2 (\mu_{c},\mu_{bg})\nonumber 
\end{align}
where $N_c$ is the number of pixels in cluster $c$, $\mu_c$ is its mean embedding, and $x_i$ is the embedding of pixel $i$ within this cluster. $N_{bg}$ is the number of background pixels. $\alpha$ is the weight to balance foreground and background clusters and $\gamma$ is the weight for the regularization term. The total loss for the IEN training is the average of the per-image loss on the training set.

\vspace{-0.4cm}

%% file: tex/exp.tex
\section{Experiment}
\label{sec:exp}
We evaluate our method on two public benchmarks, i.e., MultiOrgan~\cite{kumar2017dataset} and TNBC~\cite{naylor2018segmentation}, and compare with several state-of-the-art methods~\cite{qu2019weakly,yoo2019pseudoedgenet,tian2020weakly,dong2020towards}. 
Below we first introduce the datasets and metrics, then present our empirical results comparing to other methods, and finally conduct ablation study to demonstrate the effectiveness of our design.

\begin{table}[t!]
	\caption{Ten-fold cross-validation on MultiOrgan and TNBC.}
	\label{tab:results}
	\vspace{-0.3cm}
	
	\begin{center}
		\resizebox{0.48\textwidth}{!}{
			\begin{tabular}{lccccccccc}
				\toprule
				Method & IoU & F1 & Dice & AJI \\
				\midrule
				MultiOrgan\\
				\midrule
				\cite{yoo2019pseudoedgenet} &0.6136$\pm$0.04&-&-&-\\
				\cite{qu2019weakly} &0.5789$\pm$0.06&0.7320$\pm$0.05&0.7021$\pm$0.04&0.4964$\pm$0.06\\ 
				\cite{tian2020weakly} &0.6239$\pm$0.03&0.7638$\pm$0.02&0.7132$\pm$0.02&0.4927$\pm$0.04\\
				Ours &$\mathbf{0.6494\pm0.02}$&$\mathbf{0.7863\pm0.02}$&$\mathbf{0.7394\pm0.02}$&$\mathbf{0.5430\pm0.04}$ \\
				\midrule
				TNBC\\
				\midrule
				\cite{yoo2019pseudoedgenet} &0.6038$\pm$0.03&-&-&-\\
				\cite{qu2019weakly} &0.5420$\pm$0.04&0.7008$\pm$0.04&0.6931$\pm$0.04&0.5181$\pm$0.05\\
				\cite{tian2020weakly} &$\mathbf{0.6393\pm0.03}$&0.7510$\pm$0.04&0.7413$\pm$0.03&0.5509$\pm$0.04\\
				Ours &$0.6153\pm0.03$&$\mathbf{0.7600\pm0.02}$&$\mathbf{0.7492\pm0.02}$&$\mathbf{0.5854\pm0.03}$ \\
				\bottomrule
			\end{tabular}
		}
	\end{center}
\end{table}

\begin{figure}[t!]
	\centering
	\centerline{\includegraphics[width=0.48\textwidth]{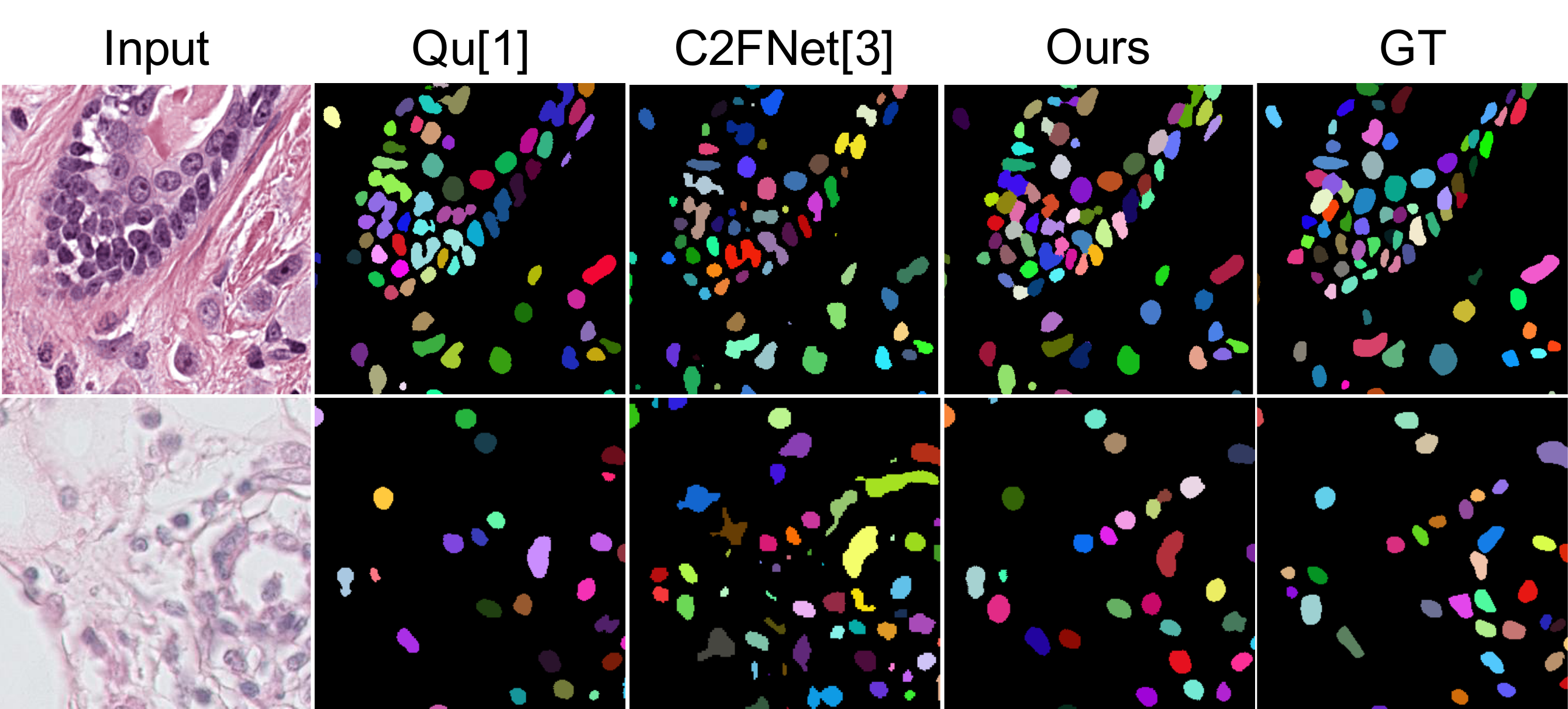}}
	%
	%
	\vspace{-0.3cm}
	\caption{Qualitative results on two datasets. We use distinct colors to denote different nuclei. Top row: MultiOrgan; Bottom row: TNBC. Our method can separate connected nuclei, as well as produce better semantic foreground.}
	\label{fig:vis}
	%
\end{figure}

\subsection{Datasets and Metrics}
We conduct experiments on two public nuclei segmentation datasets MultiOrgan~\cite{kumar2017dataset} 
and TNBC~\cite{naylor2018segmentation}. Both datasets are H\&E stained histopathology images with pixel-level masks, and we adopt the point annotation from~\cite{qu2019weakly}, which is the center of the bounding box of each nucleus mask. MultiOrgan contains 30 images of size $1000\times 1000$ from multiple hospitals and different organs, including about 21k annotated nuclei with large appearance variations. TNBC consists of 50 images of size $512\times 512$ from Triple Negative Breast Cancer (TNBC) patients, containing about 4k annotated nuclei. 

Our main results and comparisons are shown in a ten-fold cross-validation setting as adopted by~\cite{tian2020weakly}. To conduct ablation study and additional comparison with~\cite{dong2020towards}, we also show results on MultiOrgan official splits, with 12 images for training, 4 for validation, and 14 for test as the final results. 
Following~\cite{tian2020weakly}, we take both pixel-level (IoU and F1 score) and object-level (Dice coefficient~\cite{sirinukunwattana2015stochastic} and Aggregated Jaccard Index (AJI)~\cite{kumar2017dataset}) metrics for evaluation.

\subsection{Implementation details}
We adopt the same data augmentation operations from~\cite{qu2019weakly}. 
The dimension $D$ of instance feature $\mathbf{F}$ is 16. 
For k-means clustering pseudo label generation for MultiOrgan, global scaling term $\lambda$, patch size $L$, and maximum distance $d^*$ are 0.2, 60, and 20, respectively. For TNBC, the corresponding hyperparameters are 0.12, 80, and 18. 
For instance pseudo label generation, we set $T_{fg}$ as 0.5 for both datasets, and $T_{bg}$ as 0.3 for MultiOrgan and 0.2 for TNBC. 
For training of IEN, $\delta_{intra}$ and $\delta_{inter}$ are set as 0.5 and 3, respectively, and $\gamma$ is set as 0.001. 
Additionally, $\alpha$ is 1 and 0.5 for MultiOrgan and TNBC, respectively. 
We initialize the ResNet34~\cite{he2016deep} encoders of SPN and IEN with pretrained weights on ImageNet, as in~\cite{qu2019weakly}.  
We use the Adam optimizer. 
On both datasets, we train SPN with a learning rate of 1e-4 for 60 epochs, and then train IEN with 1e-3 for 1000 epochs. 
For inference, probability threshold $T_0$ is 0.3 and the bandwidth of mean-shift is 1.5 for both datasets. 
The training of our SPN and IEN on each dataset takes 1 hour and 12 hours on a single TITAN Xp GPU card, respectively. The inference of our full model on each input image takes around 6 seconds, of which 99\% is for mean-shift clustering.

\subsection{Results}
We present our model performance in ten-fold cross-validation in Table~\ref{tab:results}, and results for other methods are from~\cite{tian2020weakly}. 
On MultiOrgan, our method significantly outperforms previous works in all metrics. 
We achieve a performance gain of 7.05\% in IoU and 4.84\% in AJI, compared to our baseline method~\cite{qu2019weakly}. 
Additionally, we also outperform~\cite{tian2020weakly} by 2.55\% in IoU and 5.03\% in AJI. 
This shows the efficacy of our design in decoupling and fusing semantic and instance segmentation process, as well as the effectiveness of our learned instance-aware representation. 
As shown in Fig.~\ref{fig:vis}, our model is able to split connected nuclei and to produce better semantic foreground. 
On TNBC, our method also outperforms~\cite{qu2019weakly} with a large margin, and we achieve better results than~\cite{tian2020weakly} in F1, Dice, and especially AJI with 3.45\%. 
Additionally, our method (SPN+IEN) also outperforms~\cite{dong2020towards} in all metrics with considerable margins, as in Table~\ref{tab:ablation}.

\subsection{Ablation Study}
To illustrate the efficacy of our design, we conduct ablation study on MultiOrgan train-val-test splits, as shown in Table.~\ref{tab:ablation}. \textbf{Baseline} is the SPN trained on semantic pseudo labels generated from original k-means clustering and Voronoi partition strategy in~\cite{qu2019weakly}. \textbf{SPN} shows the improvement over \textbf{Baseline} with our adaptive k-means clustering design, which is 1.93\% in IoU and 1.49\% in F1. This shows that with our pseudo labels, \textbf{SPN} can generate better semantic proposals for IEN. Our final model \textbf{SPN+IEN} further improves Dice by 2.64\% and AJI by 6.03\% over \textbf{SPN}, demonstrating the efficacy of our instance representation learning. Additionally, \textbf{SPN+IEN} also improves pixel-level metrics, which shows that our fusion of semantic and instance segmentation is effective. 
Moreover, we show the results of our fully supervised model as \textbf{Fully}, where we use ground truth masks to train the SPN and IEN, and keep the same inference procedure as our method. The performance gaps between \textbf{Fully} and \textbf{SPN+IEN} are only around 1\% on all metrics, demonstrating the efficacy of our method in utilizing weak labels.

\begin{table}[t!]
	\caption{Comparison on the test set of MultiOrgan.}
	\label{tab:ablation}
	\begin{center}
		\resizebox{0.4\textwidth}{!}{
			\begin{tabular}{lccccccccc}
				\toprule
				Method & IoU & F1 & Dice & AJI \\
				\midrule
				Fully & 0.6701 & 0.8007 & 0.7548 & 0.5492 \\
				\midrule
				\cite{dong2020towards} &-&0.7821&0.7237&0.5202\\
				Baseline &0.6379&0.7768&0.7342&0.5263\\
				SPN &0.6572&0.7917&0.7173&0.4754\\
				SPN+IEN &$\mathbf{0.6642}$&$\mathbf{0.7967}$&$\mathbf{0.7437}$&$\mathbf{0.5357}$ \\
				\bottomrule
			\end{tabular}
		}
	\end{center}
\end{table}

%% file: tex/conclusion.tex
\section{Conclusion}
\label{sec:conclude}
In this paper, we have developed a novel framework for weakly supervised nuclei segmentation via an instance-aware learning strategy. 
Our modular deep neural network is able to decouple semantic and instance segmentation and then to perform fusion for these two subtasks. 
With the decoupling design, our model learns expressive instance representations which can be effectively used for grouping. 
Empirical results on two public benchmarks demonstrate that our method is robust for datasets with nuclei from different organs, and consistently outperforms previous state-of-the-art methods with considerable margins. 
For future work exploration, our method can potentially extend to scenarios of inaccurate or partial nuclei centroid point labels, to further reduce annotation cost.